\documentclass[prl,twocolumn,tightenlines,showpacs,nofootinbib,superscriptaddress]{revtex4-1}%
\usepackage{bm,dcolumn,amsmath,graphicx,amsfonts,amssymb}

\usepackage{hyperref} 
\usepackage{xcolor}
\definecolor{cite}{rgb}{0.,0.,0.85}   
\hypersetup{ 
	colorlinks,
	linkcolor={cite},
	citecolor={cite},
	urlcolor={cite}
}

\newcommand{\abs}[1]{\ensuremath{\left |#1\right |}}
\newcommand{\bra}[1]{\ensuremath{\langle #1|}}	
\newcommand{\ket}[1]{\ensuremath{|#1\rangle}}	
\newcommand{\Exp}[1]{\ensuremath{e^{#1}}}
\renewcommand{\v}[1]{\ensuremath{\boldsymbol{#1}}}		
\newcommand{\E}[1]{\ensuremath{\times10^{#1}}}	
\newcommand{\twocomp}[2]{\ensuremath{\begin{pmatrix}#1\\#2\end{pmatrix}}}	

\newcommand{\eq}[2]{
	\begin{equation}\label{#1}
		#2
	\end{equation}
}
\newcommand{\aln}[1]{
	\begin{align}
		#1
	\end{align}
}

\renewcommand{\d}{\ensuremath{{\,\rm d}}}	
\newcommand{\g}{\ensuremath{\gamma}} 
\renewcommand{\a}{\ensuremath{\alpha}} 
\newcommand{\s}{\ensuremath{\sigma}} 
\renewcommand{\k}{\ensuremath{\kappa}} 
 
\newcommand{\ec}{\ensuremath{\varepsilon}} 

\usepackage[normalem]{ulem}
\definecolor{newc}{rgb}{0.,0.,0.5}

\begin{document}

\title{Ionization of Atoms by Slow Heavy Particles, Including Dark Matter}

\author{B. M. Roberts}\email[]{b.roberts@unsw.edu.au}
	\affiliation{School of Physics, University of New South Wales, Sydney, New South Wales 2052, Australia}
\author{V. V. Flambaum} 
	\affiliation{School of Physics, University of New South Wales, Sydney, New South Wales 2052, Australia}
	\affiliation{Mainz Institute for Theoretical Physics, Johannes Gutenberg University Mainz, D 55122 Mainz, Germany}
\author{G. F. Gribakin} 
	\affiliation{School of Mathematics and Physics, Queen's University Belfast, Belfast BT7 1NN, United Kingdom}

\date{ \today }  

\begin{abstract}

Atoms and molecules can become ionized during the scattering of a slow, heavy particle off a bound electron.
Such an interaction involving leptophilic weakly interacting massive particles (WIMPs) is a promising possible explanation for the anomalous $9\sigma$ annual modulation in the DAMA dark matter direct detection experiment [R. Bernabei \textit{et al.}, \href{https://dx.doi.org/10.1140/epjc/s10052-013-2648-7}{Eur. Phys. J. C {\bf 73}, 2648 (2013)}].
We demonstrate the applicability of the Born approximation for such an interaction by showing its equivalence to the semiclassical adiabatic treatment of atomic ionization by slow-moving WIMPs.
Conventional wisdom has it that the ionization probability for such a process should be exponentially small.
We show, however, that due to nonanalytic, cusp-like behaviour of Coulomb functions close to the nucleus this suppression is removed, leading to an effective atomic structure enhancement.
We also show that electron relativistic effects 
actually give the dominant contribution to such a process, 
enhancing the differential cross section by up to 1000  times. 
\end{abstract}


\maketitle

\section{Introduction}

It is possible for an atom or a molecule to eject a bound electron and become ionized due to the scattering of a slow, heavy particle, such as weakly interacting massive particle (WIMP) dark matter. Such electrons can be detected, although the momentum transfer values involved in this process are very high on the atomic scale.
This type of scattering is particularly relevant for the analysis of the anomalous DAMA annual modulation signal.

The DAMA Collaboration uses a scintillation detector to search for possible dark matter (DM) interactions within NaI crystals (see Ref.~\cite{Bernabei2013} and references therein).
The data from the combined DAMA/LIBRA and DAMA/NaI experiments indicates an annual modulation in the event rate at around 3 keV electron-equivalent energy deposition with a 9.3$\sigma$ significance  (the low-energy threshold for DAMA is $\sim2$ keV) \cite{Bernabei2013}.
The phase of this modulation agrees very well with the assumption that the signal is due to the scattering of DM particles (e.g., WIMPs) present in the DM galactic halo.
This result stands as the only enduring DM direct-detection claim to date.

On the other hand, null results from several other, more sensitive, experiments (e.g., Refs.~\cite{Aprile2012,Akerib2014,AgneseSCDMS2014}) all but rule out the possibility that the DAMA signal is due to a WIMP--nucleus interaction. 
While the DAMA experiment is sensitive to scattering of DM particles off both electrons and nuclei, most other DM detection experiments reject pure electron events in order to search for nuclear recoils with as little background as possible.
This means that DM particles that interact favorably with electrons rather than nucleons could potentially explain the DAMA modulation without being ruled out by the other null results.
This possibility has been considered previously in the literature---see, e.g.,~Refs.~\cite{Bernabei2008a,Kopp2009}. 
%

In Ref.~\cite{Kopp2009}, a theoretical analysis concluded that due to a suppression in the WIMP--electron-scattering ionization cross section, loop-induced WIMP--nucleus scattering would dominate the relevant event-rate, even if the DM particles only interacted with leptons at tree level. In this case,  the previous constraints from nuclear recoil experiments \cite{Aprile2012,Akerib2014,AgneseSCDMS2014} still apply.
However, these conclusions are based on calculations that employed simple nonrelativistic wave functions. 
A rigorous {\em ab initio} relativistic treatment of the atomic structure has not yet been implemented, and, as we demonstrate in this work, is crucial.

A recent analysis of data from the XENON100 experiment has also investigated WIMP-induced electron-recoil events~\cite{XENONcollab2015,XENONcollab2015a}, and also observed modest evidence for an annual modulation (at the $2.8\sigma$ level).
However, based on their analysis of the average unmodulated event-rate, DM that interacts with electrons via an axial-vector coupling was excluded as an explanation for the DAMA result at the $4.4\s$ level \cite{XENONcollab2015}.
This means that for DM--electron scattering to be consistent with both the DAMA and XENON experiments, both event rates would have to have a very large modulation fraction.
It has been suggested that electron-interacting ``mirror'' DM may satisfy this criteria \cite{Foot2015} (see also Ref.~\cite{Foot2014}).
By assuming the DAMA result was due to an axial-vector coupling, and using the theoretical analysis from Ref.~\cite{Kopp2009}, the corresponding modulation amplitude that would be expected in the XENON experiment was calculated in Ref.~\cite{XENONcollab2015a}. 
The observed amplitude was smaller than this by a factor of a few, and it was concluded that the XENON results were inconsistent with the DAMA results at the $4.8\s$ level \cite{XENONcollab2015a}.
We note, however, these conclusions are not entirely model independent, and a rigorous relativistic analysis is required.
We also note that there is no {\em a priori} reason to believe that the fraction of the modulated signal should be small or proportional to the fractional annual change in the DM velocity distribution. 
In fact, the scattering amplitude is very highly dependent on the values of momentum transfer involved, which depend on the velocity of the DM particles.
As we shall show, electron relativistic effects must be taken into account properly to recover the correct momentum-transfer dependence of the cross section.

In this work, we demonstrate the applicability of the Born approximation to such scattering processes, and demonstrate that the expected exponential suppression of the relevant electron matrix elements is lifted.
This essentially amounts to an effective enhancement of the scattering rate due to the behaviour of the electron wave functions at very small distances (i.e., close to the nucleus).
Crucially, we also show that relativistic effects actually give the dominant contribution to the scattering cross section. This means that an analysis that only considers nonrelativistic electron wave functions may underestimate the size of the effect by several orders of magnitude, which may have significant implications for the interpretation of electron-recoil experiments, such as DAMA {and XENON}.
A more complete {\em ab initio} {relativistic many-body} calculation of the scattering cross section and ionization rate will be presented in a later work.

As well as the DAMA and XENON100 experiments, our results also have implications for the interpretation
of other experiments which are sensitive to electron recoils, such as XENON10 \cite{Essig2012a,*Essig2012,Angle2011} and CoGeNT \cite{CoGeNT2013,*CoGeNT2014}. They will also be useful for the planning and interpretation of future experiments, in particular for those proposed to search for light DM for which electron recoils are significant \cite{Essig2012a,*Essig2012,Essig2015}.

\section{Scattering amplitude}\label{sec:ampl}

\begin{figure}
		\includegraphics[height=0.25\textwidth]{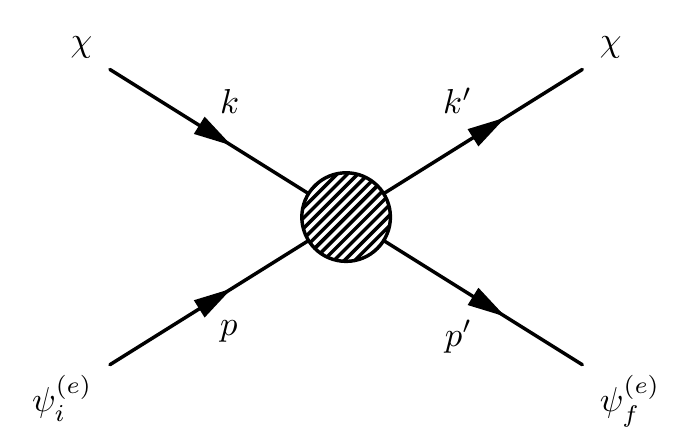}
		\caption{Effective amplitude for scattering of a heavy particle ($M_\chi\gg m_e$) off an electron, via a short-range interaction. We consider the case corresponding to atomic ionization, where $\psi_i$ is a bound electron state and $\psi_f$ is a state in the continuum.}
		\label{fig:scattering}
\end{figure}

We consider the scattering amplitude and the resulting cross section for the ionization of an atom or molecule via the scattering of a slow, heavy particle $\chi$ (e.g., a dark-matter WIMP), off the atomic electrons; see Fig.~\ref{fig:scattering}.
By heavy and slow, we mean that the mass of the incident particle is much greater than the electron mass, $M_\chi \gg m_e$, while the typical velocity of the DM particle $V$ is much smaller than the characteristic electron velocity,
$V\ll v_e$, where $v_e\sim\a c$ ($\a \approx1/137$ is the fine-structure constant and $c\approx3\E{8}\;{\rm ms}^{-1}$ is the speed of light).

For simplicity, we take the effective interaction between the particle $\chi$ and the electron to be a pure contact interaction
\eq{eq:veff}{
\hat V_{\rm eff}=\a_{\chi}\delta\left( \v{r}-\v{R} \right),
}
where $\v{r}$ and $\v{R}$ are the position vectors of the electron and the heavy particle, respectively, and $\a_\chi $ is the strength constant. 
(Considering a finite-range interaction does not affect our conclusions; see the Appendix.)
Assuming that the interaction is weak, the amplitude 
is given by the Born approximation,
\aln{
M&=\a_\chi\int \Exp{-i\v{k}'\cdot\v{R}}\psi_f^*(\v{r})\delta\left( \v{r}-\v{R} \right) \Exp{i\v{k}\cdot\v{R}}\psi_i(\v{r})\d^3\v{r}\d^3\v{R} \notag\\
&=\a_\chi\bra{f}\Exp{-i\v{q}\cdot\v{r}}\ket{i},\label{eq:M}
}
where $\v{k}$ and $\v{k}'$ are the initial and final momenta of the heavy particle, respectively, and $\v{q}=\v{k}'-\v{k}$ is the momentum transfer (we use units in which $\hbar =1$).

From Eq.~(\ref{eq:M}), the scattering cross section is expressed as
\eq{eq:dsig}{
\d\s =
\frac{2\pi}{V}\a_\chi^2\abs{\bra{f}\Exp{-i\v{q}\cdot\v{r}}\ket{i}}^2\delta\left(\frac{k^2-k'^2}{2M_\chi}-\omega\right)\frac{\d^3k'}{(2\pi)^3}\d\nu_f,
}
where $V=k/M_\chi $, $\omega=\ec_f-\ec_i$ is the change in energy between the initial and final electron states, and $d\nu_f$ is the density of final electron states. Assuming that the heavy particle is incident along the $z$ axis, we obtain from Eq.~(\ref{eq:dsig}),
\eq{eq:born-r}{
\d\s=\frac{\a_\chi^2}{V^2}\abs{\bra{f}\Exp{-i\v{q}\cdot\v{r}}\ket{i}}^2\frac{q\d q\d\phi }{(2\pi)^2}\d\nu_f
,}
where $q^2=k'^2+k^2-2k'k\cos \theta $, $\theta $ is the heavy-particle scattering (polar) angle, $\phi $ is the azimuthal angle, and $k'=\sqrt{k^2-2M_\chi \omega }$. The total cross section is obtained from Eq.~(\ref{eq:born-r}) by integration over $q$ and $\phi $ and summation/integration over the initial and final electron states.

Since the energy of the DM particle is much greater than the typical electron transition energy $\omega $, the smallest momentum transfer is $q_{\rm min}=|k'-k|\simeq \omega /V$. The corresponding vector $\v{q}$ is in the negative $z$ direction. For low incident velocities $V$ this momentum is much larger than the typical electron momentum $p_e$:
$q_{\rm min}\gtrsim m_ev_e^2/V=p_e(v_e/V)\gg p_e$. In this case the Fourier transform in the amplitude 
(\ref{eq:M}) is suppressed and the momentum transfers essential for the total cross section are such that $q\ll k$ (though formally Eq.~(\ref{eq:born-r}) is integrated up to $q_{\rm max}=k+k'$). 
Hence, we can write the momentum transfer vector as $\v{q}=-(\omega /V)\hat{\v{e}}_z+\v{q}_\perp$, where $\v{q}_\perp$ is the component perpendicular to $z$. 
This also implies $q\d q\d\phi =\d^2q_\perp $. 
Changing the electron wave functions from the coordinate to momentum space (e.g., $\widetilde \psi _i (\v{p})=\int \Exp{-i\v{p}\cdot \v{r}}\psi _i(\v{r})\d^3\v{r}$), we have from Eq.~(\ref{eq:born-r}):
\aln{
\d\s=&\frac{\a_\chi^2}{V^2}
\abs{\int\widetilde\psi_f^*(\v{p}-\v{q})\widetilde\psi_i(\v{p})\frac{{\rm d}^3p}{(2\pi)^3}}^2
\frac{{\rm d}^2q_\perp}{(2\pi)^2}\d\nu_f
\label{eq:born-q}
.
}

Let us now derive this result using a different approach, i.e., the time-dependent perturbation theory, and treating the motion of the heavy particle classically. 
We assume that this particle moves in a straight line, $\v{R}=\v{\rho }+\v{V}t$, where $\v{\rho }$ is the impact parameter vector perpendicular to $\v{V}$. The probability of the transition $i\to f$ induced by the interaction (\ref{eq:veff}) is then given by $\d w=\abs{M_{\rm td}}^2\d\nu_f$, where the transition amplitude is \cite{LLVol3}
\eq{eq:tdamp}{
M_{\rm td}=\a_\chi\int^\infty_{-\infty}\Exp{i\omega t}\bra{f}\delta\left(\v{r}-\v{\rho}-\v{V} t\right)\ket{i}\d t
.}
Replacing the electron wave functions with their momentum-space counterparts, $\widetilde \psi _i (\v{p})$ and $\widetilde \psi _f (\v{p}')$, and integrating over $\v{r}$ and $t$ gives the amplitude in the form
\eq{eq:Mtd}{
M_{\rm td}=\frac{\a_\chi}{V}
\iint
\Exp{i\v{q}_\perp\cdot\v{\rho}}\widetilde\psi_f^*(\v{p}-\v{q})\widetilde\psi_i(\v{p})\frac{{\rm d}^3p}{(2\pi)^3}\frac{{\rm d}^2q_\perp}{(2\pi)^2},
}
where we introduced $\v{q}=\v{p}-\v{p}'$, whose $z$ component (i.e., that in the direction of $\v{V}$) is fixed by integration over time and given by $q_z=-\omega /V$.

The differential cross section is obtained by integrating the transition probability over the impact parameters:
\eq{eq:semicl}{
\d\s =\int \abs{M_{\rm td}}^2 \d\nu_f\d^2\v{\rho}.
}
Inserting the amplitude from Eq.~(\ref{eq:Mtd}) immediately leads to the cross section (\ref{eq:born-q}), derived earlier using the Born approximation.

This equivalence confirms the validity of the Born approximation, ensuring that Eq.~(\ref{eq:born-r}) can be used to calculate the cross section and event-rates. 
At the same time, the time-dependent picture provides additional understanding of the smallness of the ionization cross section by WIMPs (beyond the overall factor $\a_\chi^2$). 
It is known that if the time scale of the perturbation $T$ is much larger than the characteristic period $\tau $ of the system, the effect of the perturbation is suppressed \cite{LLVol3}. 
In our case $T=a_0/V$ and $\tau \sim 1/\omega $, where $a_0$ is the Bohr radius (or, more generally, the radius of the electron orbit). 
Their ratio is $T /\tau \sim a_0\omega /V \sim q_{\rm min}/p_e \gg 1$
\cite{[{In atomic collisions this is known as the Massey adiabatic criterion; see }]Massey1949}, 
which means that the cross section is suppressed. 
This suppression is in general exponential, unless the potential in which the particles 
move has a singularity, which changes the suppression into a power law.
This is demonstrated explicitly 
 in the next sections.

\section{Exponential suppression \\(and lack thereof)}\label{sec:supp}

The large magnitude of the momentum transfer $q$ relative to the typical electron momenta means that the exponent in the atomic structure factor $\bra{f}\Exp{-i\v{q}\cdot\v{r}}\ket{i}$ oscillates much more rapidly than the electron wave functions involved. 
The value of this integral is thus determined by small electron--nucleus separations, and the dominant contribution to the cross section is proportional to the probability of finding the electron close to the nucleus.

In general, rapid oscillations of $\Exp{-i\v{q}\cdot\v{r}}$ lead to an exponential suppression of the amplitude. The simplest way to see this is by assuming that the electron wave functions have an oscillator-like behaviour, $\psi_{i,f}(r)\sim A\Exp{-\beta r^2}$.
The matrix element for large $q$ will then be 
\eq{eq:suppress}{
\bra{f}\Exp{-i\v{q}\cdot\v{r}}\ket{i}\propto \Exp{-{q^2}/{8\beta}},
}
which is exponentially suppressed. 
This behaviour will be observed for any electron wave functions that are smooth near the origin.
As mentioned 
above,
exponential suppression of the amplitude for large $v_e/V$ is also the general result of the adiabatic nature of the perturbation by a slow projectile \cite{LLVol3,Massey1949}.

We demonstrate, however, that the behaviour of the electron wave functions near the origin of the Coulomb field leads to contributions to the amplitude that are not exponentially suppressed. These terms are proportional to the nuclear charge $Z$ and decrease only as a power of $q$ at large $q$.

Consider the ejection of an electron with energy $\ec $ from an atomic orbital $nl$. The contribution of the final-electron partial wave $l'$ to the amplitude (\ref{eq:M}) is proportional to the radial integral
\eq{eq:rad}{
\int _0^\infty R_{\ec l'} (r) R_{nl}(r)j_L(qr)r^2\d r,
}
where $R_{nl}(r)$ and $R_{\ec l'}(r)$ are the radial wave functions of the initial and final states, $j_L(x)$ is the spherical Bessel function, the values of $l$, $l'$ and $L$ must satisfy the triangle inequality, and $l+l'+L$ must be even due to parity selection. The leading contribution to this integral at large $q$ comes from small $r\sim 1/q$, where the radial functions behave as (in atomic units: $a_0=1$, $c=1/\a\approx 137$)
\eq{eq:R_nl}{
R_{nl}(r)\simeq A r^{l}\left[1-\frac{Z}{l+1}r+\dots \right],
}
$A$ being the normalization factor, and with a similar expression for $R_{\ec l'}(r)$. From Eqs.~(\ref{eq:rad}) and (\ref{eq:R_nl}), it appears that the leading contribution to the amplitude at high $q$ is proportional to
\eq{eq:rad_1}{
\int _0^\infty r^{l+l'+2}j_L(qr)\d r=
\frac{1}{q^{l+l'+3}}\int _0^\infty x^{l+l'+2}j_L(x)\d x.
}
However, the integral
\eq{eq:Bes_int}{
\int _0^\infty x^{l+l'+2}j_L(x)\d x=2^{l+l'+1}\sqrt{\pi }
\frac{\Gamma\left[ \frac{3}{2}+\frac{1}{2}(L+l+l')\right]}{\Gamma\left[ \frac{1}{2}(L-l-l')\right]}\notag
}
is identically zero for even $l+l'+L$, since the gamma function in the denominator has poles for nonpositive integer arguments. This also shows that including any even-power corrections to the small-$r$ expansion of the wave functions, Eq.~(\ref{eq:R_nl}) (which appear for any potential regular at the origin), also leads to zero contributions. Therefore the amplitude in this case would decrease faster than any power of $q$, i.e., exponentially.

On the other hand, the lowest-order, linear correction in either $R_{nl}(r)$ and $R_{\ec l'}(r)$, is proportional to $Z$ [see Eq.~(\ref{eq:R_nl})], and the integral $\int _0^\infty x^{l+l'+3}j_L(x)\d x$ is \textit{nonzero}. This determines the leading asymptotic behaviour of the amplitude,
\eq{eq:rad2}{
\int_0^\infty R_{\ec l'} (r) R_{nl}(r)j_L(qr)r^2\d r\propto \frac{Z}{q^{l+l'+4}}.
}
This shows that the largest cross section for large $q$ (i.e., the least suppression, 
$\abs{\bra{f}\Exp{-i\v{q}\cdot\v{r}}\ket{i}}^2\propto q^{-8}$) 
is obtained for both initial and final $s$ states ($l=l'=L=0$).
{Similar integrals arise in the problem of atomic photoionization at high energies, see, e.g., Ref.~\cite{Dalgarno1992,Forrey1995}.}


This power, instead of the exponent, emerges due to the Coulomb singularity of the electron wave function at the nucleus.
The singularity for the $s$-wave electrons is stronger than in higher partial waves, whose expansions contain extra powers of $r$. 
The ionization by slow, heavy particles is thus dominated by $s$-wave contributions from small electron--nuclear distances. 
This means that there is an effective atomic-structure enhancement of such scattering processes involving $s$-waves, and that relativistic effects may be significantly larger than expected.

\section{Relativistic Enhancement}\label{sec:rel}

Indeed, the situation in the relativistic case is quite different.
Consider the Dirac wave function for an electron in the central field of the atom,
\eq{eq:diracwf}{
\psi_{n\k m}=
\twocomp{F_{n\k}(r)\,\Omega_{\k m}(\theta_r,\phi_r)}{i G_{n\k}(r)\, \Omega_{-\k, m}(\theta_r,\phi_r)},
}
where $\k$ is the Dirac quantum number [$\k=-(l+1)$ for $j=l+1/2$, and
$\kappa =l$ for $j=l+1/2$, $j$ being the total angular momentum], and $\Omega_{\k m}$ is the two-component spherical spinor. At small $r$, the radial functions of the large and small Dirac components behave as \cite{LLvol4} 
\begin{subequations}
\aln{
F_{n\k}(r)&\simeq B \, r^{\g-1}   \,(\g-\k+Cr+\dots ), \label{eq:diracF}
\\
G_{n\k}(r)&\simeq -Z\alpha B \, r^{\g-1}(1+Dr+\dots ), \label{eq:diracG}
}
\end{subequations}
where $B$ is a normalization constant, $\g=\sqrt{\k^2-(Z\a)^2}$,
\eq{eq:C}{
C=-\frac{Z}{2\g +1}[1+(2\g -2\k +1)(1+\a^2\ec _{n\k})],
}
$\ec _{n\k}$ is the electron energy, and 
$D\sim C$.

In the nonrelativistic limit ($Z\a \rightarrow 0$, $\g =|\k|$) $G_{n\k}$ vanishes, and Eq.~(\ref{eq:diracF}) reduces to Eq.~(\ref{eq:R_nl}).
However, the corrections in $\g=|\k |-(Z\a)^2/2|\k|+\dots $ actually change the power of $r$ that appears in the expansion. 
As a result, the lowest-order in $r$ term, which vanished in the nonrelativistic case [see Eq.~(\ref{eq:rad_1})], now becomes
\eq{}{
\int _0^\infty r^{\g+\g'}j_L(qr)\d r
=\frac{2^{\g+\g'-1}}{q^{\g+\g'+1}}\sqrt{\pi }
\frac{\Gamma\left[\frac{1}{2}(L+\g+\g'+1)\right]}{\Gamma\left[ \frac{1}{2}(L-\g-\g'+2)\right]},\notag
}
which is {\em nonzero}.
For example, for the initial and final $s$-wave states [$\k=-1$, $\g=\g '\simeq 1-(Z\a)^2/2$], we have
\eq{}{
\int _0^\infty r^{2\g }j_0(qr)\d r=\frac{\Gamma(2\g)\sin[\pi(1-\g)]}{q^{2\g+1}}
\simeq \frac{\pi(Z\a)^2}{2 \, q^{3-(Z\a)^2}}. 
}
The power of $q$ remains the same for scalar, pseudoscalar, vector, and pseudovector interactions, and if $p$ states are considered; see the Appendix. 


Thus we see that not only is the exponential suppression removed, but even the power suppression is significantly weaker than that found in the nonrelativistic case.
Note that the cross section goes as the square of the amplitude, meaning that the momentum-transfer dependence of the leading atomic structure contribution to the cross section is proportional to $q^{-6+2(Z\a)^2}$ 
(compared to $q^{-8}$ in the nonrelativistic case, and $\Exp{-(q/p_e)^2}$ in the ``naive'' adiabatic case; see also the Appendix).
This result indicates that relativistic effects actually give the dominating contribution to the amplitude, and that therefore a nonrelativistic treatment of such problems can greatly underestimate the size of the effect.

\begin{figure}
		\includegraphics[width=0.425\textwidth]{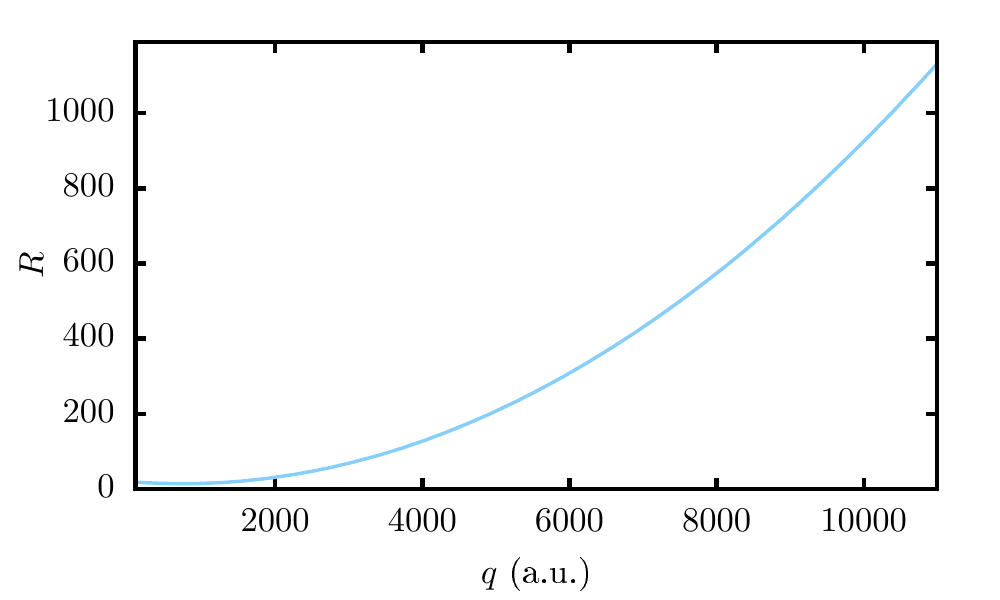}
		\caption{The ratio $R$ of the atomic structure factor 
		$\abs{\bra{\ec s_{1/2}}\Exp{-i\v{q}\cdot\v{r}}\ket{3s_{1/2}}}^2$ 
for the ionization of $3s_{1/2}$ electrons in iodine ($I_{3s_{1/2}}=1.1$~keV, $\ec =2$~keV) calculated with relativistic (hydrogen-like Dirac) wave functions to that obtained with nonrelativistic wave functions, using the true nuclear charge $Z=53$.
(The momentum conversion is $1~\mbox{MeV}/c \approx 268$ a.u.)}\label{fig:enhancement}
\end{figure}

We have verified this enhancement numerically.
In Fig.~\ref{fig:enhancement}, we plot the ratio of the atomic structure factor  calculated using relativistic Dirac-Coulomb wave functions to those calculated using nonrelativistic Schr\"odinger functions.
The calculations were performed for the $3s$ state of iodine, which would give the dominant contribution to the ionization rate due to WIMP--electron scattering at the energy scale relevant to the DAMA experiment \cite{Kopp2009} {(the $1s$ and $2s$ electrons are bound too tightly in I and Xe to become ionized with the energy deposition observed by DAMA)}.
We use the true nuclear charge ($Z=53$ for iodine) instead of the effective nuclear charge $\widetilde Z$.
The standard definition $\widetilde Z=n\sqrt{2 I_{nl}}$, where $n$ is the principal quantum number and $I_{nl}$ is the electron ionization energy, provides a sensible approximation for the wave function at intermediate distances. 
At very small distances, however, the nuclear charge is essentially unscreened. Therefore, the atomic wave functions at small distances are proportional to the pure Coulomb wave functions.
Given the importance of small electron--nuclear separations, such processes are strongly dependent on the atomic number, and using a too-small $\widetilde Z$ can lead to orders-of-magnitude underestimation of the probability.
{The best way to treat this problem is to use a self-consistent field approach, such as the relativistic Hartree-Fock method.}

\section{Conclusion}

We have demonstrated the equivalence of the first Born approximation and semiclassical treatment for the problem of ionization of atoms or molecules by slow, heavy particles, such as WIMPs, and have shown that relativistic effects (which were previously neglected) give the dominant contribution to the event rate.
Our calculations have implications for the interpretation of the anomalous annual modulation result of the DAMA and XENON collaborations in terms of DM particles scattering off the atomic electrons. 

One might suspect that this is a less-likely explanation, due to the supposed {large} suppression from the electron wave function at high momentum. 
We show, however, that this suppression is not as strong {as expected}, leading to an effective enhancement coming from the contribution of $s$-wave electrons at very small distances. 
The effect is thus highly-dependent on the nuclear charge, which is mostly unscreened at such small distances. Therefore, there are implications for simple calculations that employ screened hydrogen-like functions with an effective nuclear charge $\widetilde Z$; it is more appropriate to employ a fully self-consistent approach, such as the relativistic Hartree-Fock method. 
Crucially, relativistic corrections to such a process are significantly larger than may have been expected due to the different radial dependence of the Dirac wave functions (compared to the Schr\"odinger wave functions) at small distances.
A nonrelativistic treatment of such problems can therefore greatly underestimate the size of the effect.
Several new experiments designed to test the DAMA results are currently under way \cite{Bernabei2015,Amare2015b,*Amare2015,*Amare2015a,Xu2015}; it is crucial that the relevant theory required for their interpretation is correct.
There are also implications for the planning and interpretation of future experiments, in particular for those proposed to search for light DM, for which electron recoils are significant \cite{Essig2012a,*Essig2012,Essig2015}.

\acknowledgements
\paragraph*{Acknowledgements---}
This work was supported in part by the Australian Research Council.
We would like to thank J.~C.~Berengut, M.~Pospelov, and Y.~V.~Stadnik for helpful discussions,
and M.~Schumann for his comments on the manuscript.
BMR and VVF are grateful to the Mainz Institute for Theoretical Physics (MITP), where part of this work was completed, for its hospitality and support.


\appendix
\section{Appendix}

\paragraph{Lorentz structures---}
The lowest-order in $r$ term, which vanished in the nonrelativistic case---see Eq.~(\ref{eq:rad_1})---now becomes
\eq{eq:supp1}{
\int _0^\infty r^{\g+\g'}j_L(qr)\d r
=\frac{2^{\g+\g'-1}}{q^{\g+\g'+1}}\sqrt{\pi }
\frac{\Gamma\left[\frac{1}{2}(L+\g+\g'+1)\right]}{\Gamma\left[ \frac{1}{2}(L-\g-\g'+2)\right]},
}
which is {\em nonzero}.
For example, for the initial and final $s$-wave states [$\k=-1$, $\g=\g '\simeq 1-(Z\a)^2/2$], we have
\eq{eq:supp2}{
\int _0^\infty r^{2\g }j_0(qr)\d r=\frac{\Gamma(2\g)\sin[\pi(1-\g)]}{q^{2\g+1}}
\simeq \frac{\pi(Z\a)^2}{2 \, q^{3-(Z\a)^2}}. 
}
If one considers the contribution from a $p_{1/2}$ state ($\k =1$) for either the bound or continuum electron, or both, the power of the $q$ remains the same, but the coefficient is further suppressed by a power of $Z\a$ coming from the small component $G$.
The small component ($G$) is suppressed by the small factor $\sim Z\a$, while for $p_{1/2}$ and other $j=l-1/2$ orbitals the lowest-order term in the large component is suppressed by the factor $\k-\g\sim(Z\a)^2$.

Our results hold for scalar, pseudoscalar, vector, and pseudovector interactions. 
For the scalar, pseudoscalar, and the zero-components of the vector and pseudovector interactions, the atomic structure factors are proportional to integrals of the form 
%
\begin{align}
\abs{\bra{\ec\k'}{\Exp{i\v{q}\cdot\v{r}}\g^0}\ket{n\k}}^2
& ={C}\left(R_{ff}^2-2R_{ff}R_{gg}+R_{gg}^2\right),\\
\label{eq:rint-PS}
\abs{\bra{\ec\k'}{\Exp{i\v{q}\cdot\v{r}}\g^0\g_5}\ket{n\k}}^2
& ={D}\left(R_{fg}^2+2R_{fg}R_{gf}+R_{gf}^2\right),\\
\abs{\bra{\ec\k'}{\Exp{i\v{q}\cdot\v{r}}}\ket{n\k}}^2
&
={C}\left(R_{ff}^2+2R_{ff}R_{gg}+R_{gg}^2\right),\\
\label{eq:rint-PV}
\abs{\bra{\ec\k'}{\Exp{i\v{q}\cdot\v{r}}\g_5}\ket{n\k}}^2
& ={D}\left(R_{fg}^2-2R_{fg}R_{gf}+R_{gf}^2\right),
\end{align} 
where $C$ and $D$ are angular coeficients, and the radial integrals are
\begin{align}
R_{ff}&=\int F_{\ec\k'}F_{n\k}j_L(qr)r^2\d r,\\
R_{gg}&=\int G_{\ec\k'}G_{n\k}j_L(qr)r^2\d r,\\
R_{fg}&=\int F_{\ec\k'}G_{n\k}j_L(qr)r^2\d r,\\
R_{gf}&=\int G_{\ec\k'}F_{n\k}j_L(qr)r^2\d r,
\end{align}
respectively (note that, for simplicity, we assume a single-particle picture for the electron matrix element, with the initial and final electron states $\psi _i$ and $\psi _f$ calculated, e.g., in the relativistic Hartree-Fock approximation. In general, one can consider many-electron wave functions, or simply sum over the single-particle states of the atomic electrons when calculating the cross section).
Our conclusions (that the relativistic electron effects dominate) hold in all of these cases.
Our result is therefore model independent; model dependencies come into the problem only at the stage of the entire DM scattering cross section. For any model, there will be a huge relativistic enhancement so long as the momentum transfer is large on the atomic scale.

The Factor $(Z\a)^2$ in the numerator of Eq.~(\ref{eq:supp2}) comes from the expansion of the gamma function in the denominator of Eq.~(\ref{eq:supp1}), which approaches infinity as $\g$ approaches unity for $L=0$.
For the case where $L=1$, however, this denominator is finite even in the $Z\a\to0$ limit. 
Considering now the pseudoscalar and pseudovector cases (where the parity selection rule is different), with an initial (bound) $s$-state, there appears a contribution that comes from the final $s_{1/2}$ continuum state with $L=1$. 
In this case, the $(Z\a)^2$ suppression from Eq.~(\ref{eq:supp2}) is removed, instead it is replaced by just a $\sim Z\a$ suppression which comes from the small Dirac component that appears in the radial integral for the pseudoscalar case (\ref{eq:rint-PS}). 
There is another enhancement by a factor of $\sim4$ due to the few roughly equal terms in Eq.~(\ref{eq:rint-PS}).
In the zero-component pseudovector case, on the other hand, this situation does not lead to an enhancement. Instead there is huge suppression, which comes from the very large cancellation of terms in Eq.~(\ref{eq:rint-PV}).
This means that calculations of the electron structure for the pseudovector case are dominated by the spatial component terms, which were negligible for the vector case.
In this case, which we have not written down explicitly, the radial integrals reduce to the same as the zero-component of the vector case, but with different angular coefficients. The large relativistic enhancement survives.

\paragraph{Finite-range interaction---}
For a finite-range interaction, the cross section, which is to be integrated over $q$, contains the squared propagator $(q^2+m_p^2c^2)^{-2}$ ($m_p$ is the mass of the exchange particle); the atomic structure factor is unchanged, and our conclusions remain the same.

\bibliography{../../AllReferences/library}


\end{document}